\documentclass[twocolumn, prl, showpacs, aps, superscriptaddress]{revtex4-1}

\usepackage{graphicx}
\usepackage{ae,aecompl}
\usepackage{bm}
\usepackage{textcomp}
\usepackage{xcolor}

\usepackage{amssymb,amsmath}
\usepackage{wasysym}
\usepackage{hyperref}

\renewcommand{\phi}{\varphi}
\newcommand{\ket}[1]{\ensuremath{\mid #1 \rangle}}

\newcommand{\op}[1]{\ensuremath{\hat{#1}}}   % ist der Operator
\newcommand{\hbareff}{\ensuremath{\hbar_{\text{eff}} } }

\begin{document}

%%%%%%%%%%%%
%Header
%%%%%%%%%%%%

\date{\today}

\title{Observing the emergence of chaos in a many-particle quantum system}

%%%%%%%%%%%%% AUTHORS%%%%%
\author{J. Tomkovi\v{c}}

\affiliation{Kirchhoff-Institut f\"ur Physik, Universit\"at Heidelberg, Im Neuenheimer Feld 227, 69120 Heidelberg, Germany}
\affiliation{Max-Planck-Institut f\"{u}r Physik komplexer Systeme, N\"{o}thnitzer Stra{\ss}e 38, 01187 Dresden, Germany}

\author{W. Muessel}
\affiliation{Kirchhoff-Institut f\"ur Physik, Universit\"at Heidelberg, Im Neuenheimer Feld 227, 69120 Heidelberg, Germany}

\author{H. Strobel}
\affiliation{Kirchhoff-Institut f\"ur Physik, Universit\"at Heidelberg, Im Neuenheimer Feld 227, 69120 Heidelberg, Germany}
\author{S. L\"{o}ck}
\affiliation{Technische Universit\"{a}t Dresden, Institut f\"{u}r Theoretische Physik and Center for Dynamics, 01062 Dresden, Germany}
 \affiliation{OncoRay, Technische  Universit\"{a}t Dresden, Helmholtz-Zentrum Dresden-Rossendorf, Dresden, Germany}
\author{P. Schlagheck}
\affiliation{D\'{e}partement de Physique, Universit\'{e} de Li\`{e}ge, 4000 Li\`{e}ge, Belgium}

\author{R. Ketzmerick}
\affiliation{Max-Planck-Institut f\"{u}r Physik komplexer Systeme, N\"{o}thnitzer Stra{\ss}e 38, 01187 Dresden, Germany}
\affiliation{Technische Universit\"{a}t Dresden, Institut f\"{u}r Theoretische Physik and Center for Dynamics, 01062 Dresden, Germany}

\author{M.~K. Oberthaler}
\email{PoincareBirkhoff@matterwave.de}
\affiliation{Kirchhoff-Institut f\"ur Physik, Universit\"at Heidelberg, Im Neuenheimer Feld 227, 69120 Heidelberg, Germany}


%%%%%%%%%%%%
%Abstract
%%%%%%%%%%%%
\begin{abstract}

Accessing the connection between classical chaos and quantum many-body systems has been a long-standing experimental challenge. Here, we investigate the onset of chaos in periodically driven  two-component Bose-Einstein condensates, whose small quantum uncertainties allow for exploring the phase space with high resolution. By analyzing the uncertainties of time-evolved many-body states, we find signatures of elliptic and hyperbolic periodic orbits generated according to the Poincar\'{e}-Birkhoff theorem, and the formation of a chaotic region at increasing driving strengths. The employed fluctuation analysis allows for probing the phase-space structure by use of only short-time quantum dynamics.
\end{abstract}
\maketitle 

%%%%%%%%%%%%
%Main text
%%%%%%%%%%%%
The  Poincar\'{e}-Birkhoff scenario is generally recognized as a key mechanism for the transition from integrability to chaos in Hamiltonian systems \cite{Birkhoff1927,Ott2002}. It makes statements about the emergence of stable and unstable periodic orbits in the phase space of non-integrable Hamiltonian systems.
For a driven anharmonic oscillator, it implies that a resonant periodic orbit of the undriven system survives only for a pair (or more generally several pairs) of phase relations between the system and the driving. One of these periodic orbits is stable (more precisely elliptic), which implies that a small initial deviation will result in quasi-periodic regular motion in its neighborhood. The other one is unstable (hyperbolic), which leads to an exponential growth of small deviations from this periodic orbit. This gives rise to chaotic dynamics in the driven system. At stronger driving, more and more near-resonant orbits will be affected by this resonance and turn into quasi-periodic or chaotic orbits, eventually leading to an appreciably large chaotic region in the phase space of the oscillator. The controlled generation of such mixed phase spaces is a pathway to the creation of many-particle entangled states at the crossover region between chaotic sea and stable islands~\cite{Weiss2008, Xie2005,Ghose2008}, as well as chaos-assisted tunneling between stable islands~\cite{Lin1990a,Tomsovic1994,Korsch2008CAT}. The experimental observation of the Poincar\'{e}-Birkhoff scenario in many-body systems has so far been obscured by the large extension of the employed quantum states with respect to the phase-space structure~\cite{Moore1994, Moore1995,Steck2001,Hensinger2001, Chaudhury2009}.\\
%%%%%%%%%%%%% FIG1%%%%%%%%%%%%%%%
\begin{figure*}[ht]
\includegraphics[scale=1]{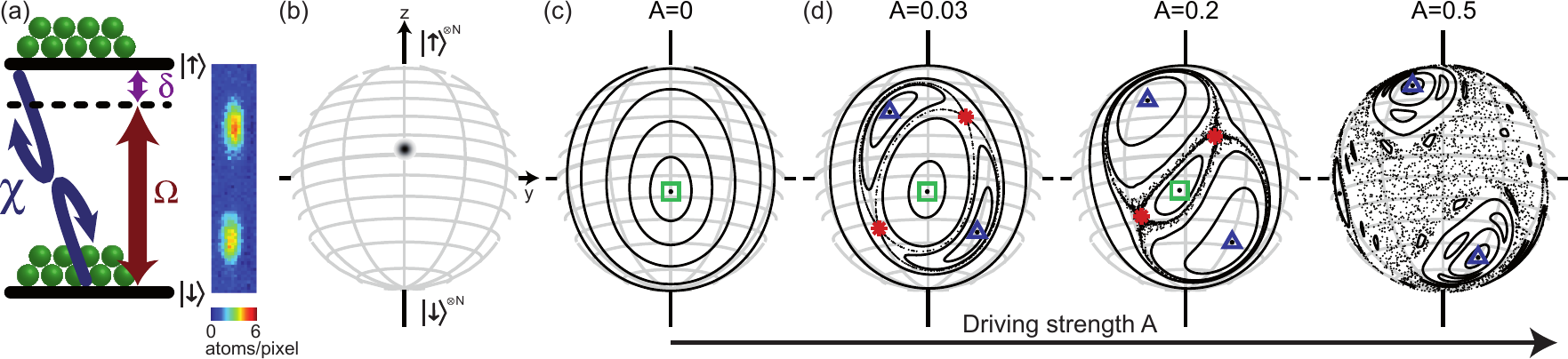}
    \caption{(Color online). {Poincar\'{e}-Birkhoff scenario in a many-body quantum system. }
  (a) A two-component Bose-Einstein condensate ($N\approx700$ atoms) with interspecies interaction $\chi$ and linear coupling $\Omega$ realizes an anharmonic oscillator with the population imbalance $z$ and the relative phase $\phi$ as the canonical variables. State selective absorption imaging allows for the precise determination of the populations  of the two components. (b) The minimal uncertainty state is a coherent spin state whose extension in the phase-space representation $(z,\phi)$ corresponds to $\hbar_{\text{eff}}=2/N$. (c) Classical orbits for the undriven case ($A=0$)  at $\Lambda=N \chi/\Omega_0 =0.7$ (green square: elliptic fixed point at $\phi=\pi$). (d) Stroboscopic Poincar\'{e} sections, showing the formation of an alternating sequence of elliptic (blue triangles) and hyperbolic  (red dots)  periodic orbits according to the Poincar\'{e}-Birkhoff theorem in the case of weak driving ($A=0.03$), the emergence of a chaotic layer for  $A = 0.2$, and a larger chaotic region at even higher modulation amplitude.}
\end{figure*}
%%%%%%%%%%%%% FIG1%%%%%%%%%%%%%%%%%
%%%%%%%%%%%%%%%%%%%%%%%%%%%%%%%%
%%%%%%%%%%%%% FIG2%%%%%%%%%%%%%%%%%
\begin{figure*}[ht]
    {\includegraphics[scale=0.9]{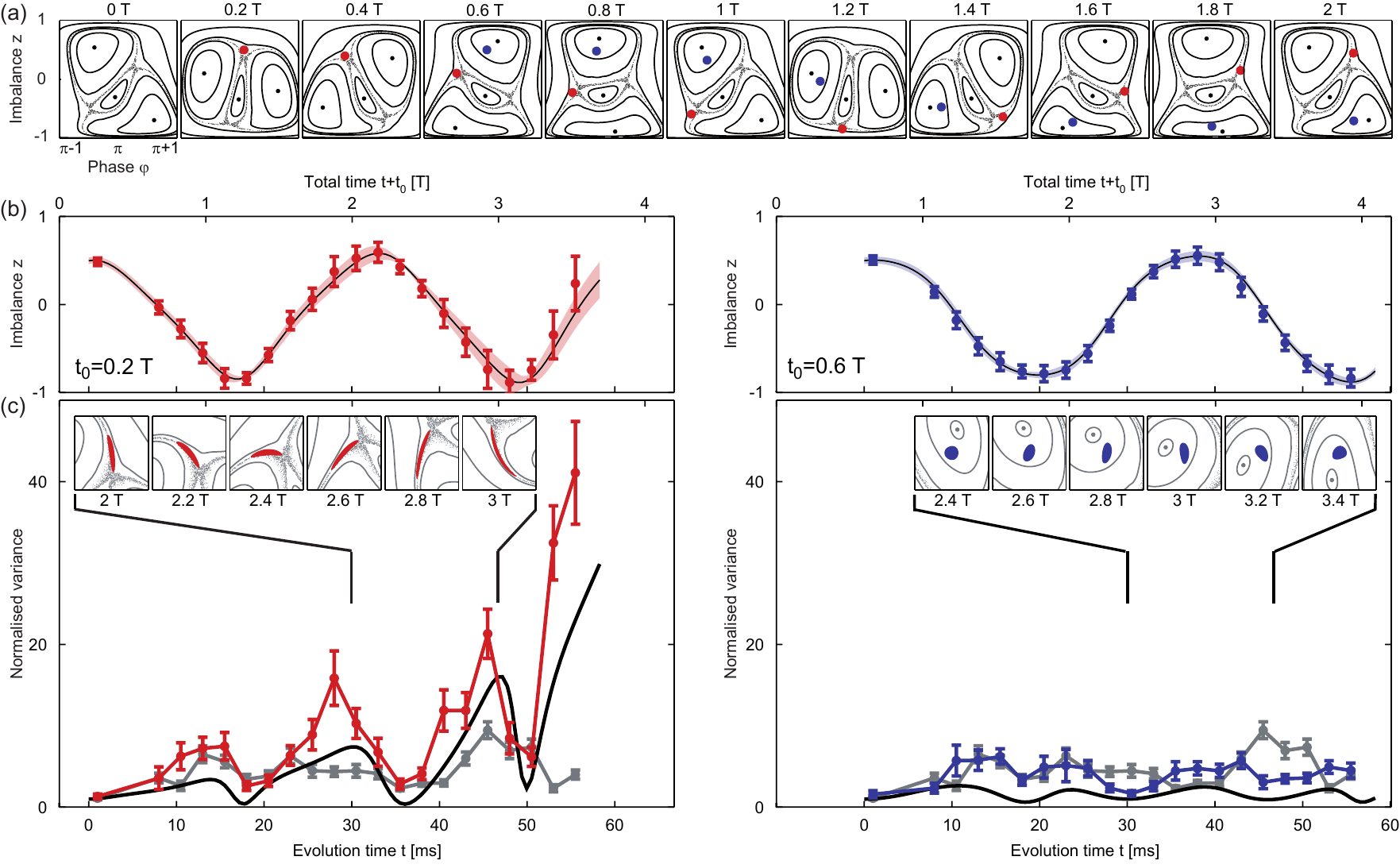}}
    \caption{(Color online). {Detecting the Poincar\'{e}-Birkhoff scenario via variance analysis.}
    (a) Stroboscopic Poincar\'{e} sections for $\Lambda=0.7$ and $A=0.2$ for various times $t \in [0, 2T]$, showing one full counterclockwise rotation of the elliptic and hyperbolic periodic orbits around the central elliptic fixed point. Red and blue points: Time evolution for $(z_0=0.55, \phi_0=\pi)$ and driving time offsets $t_0=0.2 T$ (red) and $t_0=0.6 T$ (blue). (b) The experimentally obtained mean values of $z$ exhibit oscillations with period $2T$ for preparations close to the hyperbolic orbit (left, red) and the elliptic orbit (right, blue).  Error bars are the RMS deviation.  Black lines and shaded regions: Mean and RMS uncertainty for classical time evolution of an ensemble corresponding to a CSS. (c) Variances of $z$ in units of the corresponding coherent state \cite{SuppInfo}. Preparation near the hyperbolic orbit (left) results in a strong increase compared to the undriven dynamics (gray), whereas suppressed uncertainties  can be observed close to the  elliptic orbit (right, blue). The results of classical simulations are shown as black lines. The oscillations of the variance result from the rotation of the corresponding manifolds around the central elliptic fixed point (insets). }
\end{figure*}
%%%%%%%%%%%%%%%%%%%%%%%%%%%%%%%%
In this letter, we employ time-periodically driven Bose-Einstein condensates \cite{Duffy2004,Behinaein2006}, whose small quantum uncertainties allow for exploring the phase space with high resolution \cite{Frahm1985, Salmond2002}. In this many-particle quantum system, we demonstrate signatures of elliptic and hyperbolic periodic orbits generated according to the Poincare-Birkhoff theorem, and the formation of a chaotic region at increasing driving strengths. The employed analysis of the uncertainties of time-evolved many-body states allows us to probe the corresponding phase-space structure by use of only short-time quantum dynamics.

Our system of an interacting two-component Bose-Einstein condensate (BEC) with linear coupling (see Fig.\,1a) constitutes an anharmonic oscillator if the strengths of coupling $\Omega$ and interaction-induced nonlinearity $N\chi$ are comparable. The phase and amplitude controlled coupling in our system is implemented using a two-photon transition between the two hyperfine states $\ket{\downarrow} = \ket{F=1,m_F=1}$ and $\ket{\uparrow}= \ket{2,-1}$ of $^{87}$Rb, and the nonlinearity is enhanced in the vicinity of a Feshbach resonance (details in the Supplemental Material \cite{SuppInfo}).
Our experiments investigate the phase space spanned by the population imbalance $z= (N_{\uparrow}-N_{\downarrow})/N$ and the relative phase  $\phi=(\phi_\downarrow-\phi_\uparrow)$ between the two states. Here, $N_{\uparrow}$ and $N_{\downarrow}$ are the atom numbers of the individual components, which constitute the experimental observables (see Fig.\,1a), and $N\approx700$ is the total atom number. 
External dynamics, central to related proposals and experiments with driven BECs~\cite{Gardiner2000,Duffy2004,Behinaein2006, Monteiro2008}, are frozen out and the dynamics happens exclusively in the collective internal degree of freedom.
The extension of a minimal quantum uncertainty state in this system, i.e. a coherent spin state (CSS), is given by $[\op z, \op \phi] = i2/N$, which can be interpreted as realizing an effective Planck's constant $\hbar_{\text{eff}} = 2/N$  (see Fig.\,1b). While the corresponding uncertainty is large for few-state systems~\cite{Chaudhury2009}, $\hbar_{\text{eff}}$ in our mesoscopic two-mode BEC is small and allows for probing the phase space with high resolution.

For $N \rightarrow \infty$, the dynamics of our system can be described by the classical  Hamiltonian~\cite{Smerzi1997}
\begin{equation}
    H(z,\phi,t) = \frac{\Lambda}{2} z^2 - \frac{\Omega(t)}{\Omega_0} \sqrt{1-z^2}\, \cos\phi + \epsilon z  \label{eq_Hamiltonian_classical}
\end{equation}
where the nonlinearity $\Lambda = N\chi/\Omega_0$ and the detuning $\epsilon = \delta/\Omega_0$ are rescaled with the Rabi frequency $\Omega_0$ \cite{SuppInfo}. The drive is realized by  $ \Omega(t)=\Omega_0 \left( 1+A \sin\left[\omega (t+t_0)\right] \right)$.
The resulting phase space is represented on a sphere with the polar angle $\vartheta = \operatorname{arccos} z$ and the azimuthal angle $\phi$, taking into account that $z$ and $\phi$ are bounded.  In Fig.\,1d, it is visualized  for different driving amplitudes by stroboscopic Poincar\'{e} sections at times which are multiples of $T=\frac{2 \pi} \omega$.
The undriven system for $\Lambda <1$ features an elliptic fixed point near $\left(z=0, \phi=\pi\right)$ (Fig.\,1c) and anharmonic oscillations around it.
 By driving with a frequency $\omega$, a $2:1$ resonance is realized for the orbit with the unperturbed frequency $\omega/2$, corresponding to a specific amplitude of the anharmonic oscillator (example for $\omega = 1.5\Omega_0$ shown in Fig.\,1d).
According to the Poincar\'{e}-Birkhoff theorem, this implies the destruction of the resonant orbit except for two elliptic and two hyperbolic periodic orbits. The resulting stationary points in the Poincar\'{e} section are marked as blue triangles and red stars. For increasing driving, the phase space acquires more and more structure, leading to a mixed phase space characterized by regular islands and a chaotic sea (see Fig.\,1d). Similar behavior has also been theoretically investigated for spin-exchange dynamics in Bose-Einstein condensates \cite{Kronjaeger2008, Cheng2010}.\\
%%%%%%%%%%%%% %%%%%%%%%%%%%%%%%%%
%%%%%%%%%%%%% FIG3%%%%%%%%%%%%%%%%%
 \begin{figure}[ht]
    \centering
        \includegraphics[scale=1]{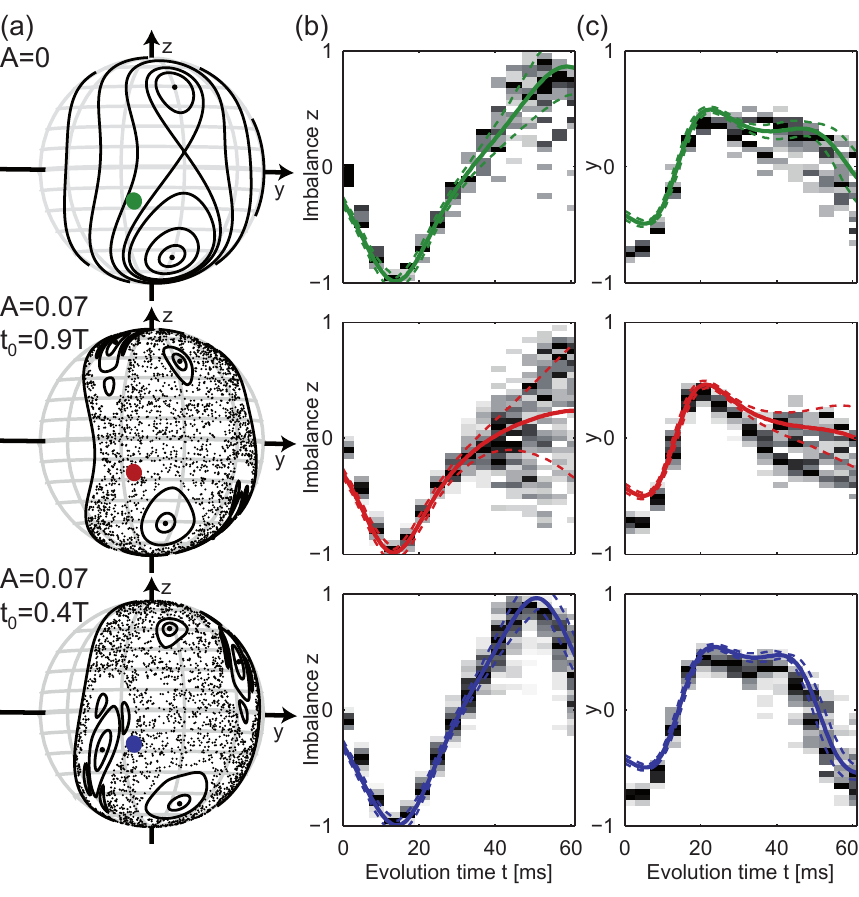}
    \caption{(Color online). {Signatures of a mixed phase space in the short-time quantum evolution.}  (a) The undriven system ($\Lambda=1.5$) exhibits one hyperbolic and two elliptic fixed points at $\phi=\pi$. For $A=0.07$, the strongly driven regime with a mixed phase space is reached, which is probed by different initial preparations (depicted by $2\sigma$ areas) with the time offsets $t_0=0.9 T$ (center, red) and $t_0=0.4 T$ (bottom, blue).  (b) Time evolution of the $z$ distributions in gray scale. Compared to the undriven case (top), we observe increased fluctuations in the case of a preparation deep in the chaotic sea (middle)  and reduced fluctuations close to a  moving island (bottom). Mean and RMS deviation of  the classical simulations are shown as solid  and dashed lines. (c) Similar signatures of the underlying phase space are observed in $y = \sin(\phi)$.}
\end{figure}
%%%%%%%%%%%%% %%%%%%%%%%%%%%%%%%%
%%%%%%%%%%%%% %%%%%%%%%%%%%%%%%%%
\indent For experimentally revealing this Poincar\'{e}-Birkhoff scenario, we prepare an initial coherent state close to either an elliptic or a  hyperbolic periodic orbit and observe the subsequent time evolution of mean and variance of the population imbalance $z$.
During the time evolution, the elliptic and hyperbolic periodic orbits perform a counterclockwise rotation around the central elliptic fixed point, as shown in Fig.\,2a. This implies identical phase spaces after one driving period $T$, and a full rotation of elliptic and hyperbolic periodic orbits after $2\,T$. All phase-space structures, e.g. regular orbits close to the elliptic periodic orbits as well as the chaotic layer, follow this rotation.\\
\indent We initialize our system with a CSS centered at $(z_0=0.55, \phi_0 =\pi)$, which is located close to  the hyperbolic periodic orbit for $t_0=0.2 \, T$ (red) and close to the elliptic periodic orbit for $t_0=0.6\,T$ (blue)  for the parameters $A=0.2$, $\Lambda = 0.7$,  $\epsilon = -0.11$ and $\omega = 1.5\Omega_0$. The mean of the quantum state (Fig.\,2b) oscillates in conjunction with the hyperbolic (red) or the elliptic orbit (blue), and resembles the dynamics of the undriven system.
However, the measured variances, corresponding to the $z$ projection of the quantum uncertainty of the time-evolved states, show striking differences, as depicted in Fig.\,2c. In classical dynamical systems, two initially adjacent points at the unstable manifold or within a chaotic region are expected to separate exponentially in time. Thus, for the preparation close to the hyperbolic orbit, the normalized variance \cite{SuppInfo} exhibits an overall increase compared to the undriven case. The observed additional oscillations result from the spreading of  the distribution along its unstable manifold, which rotates around the central elliptic fixed point  of the undriven system (inset of Fig.\,2c). Thus it constantly changes the orientation to the readout direction $z$. This behavior is well reproduced by propagating the classical Hamiltonian with an ensemble of initial conditions corresponding to a CSS  (black solid lines in Fig.\,2, \cite{SuppInfo}; quantum calculations for $N=700$ show no significant deviations). The numerical calculations underestimate the fluctuations, since experimental imperfections such as magnetic field fluctuations, particle loss and the associated decoherence and change of the system parameters are not included.\\
\indent The initial CSS close to the elliptic periodic orbit remains bounded and even shows a smaller variance compared to the undriven case (Fig.\,2c right).
This observation builds on the size of a Planck cell in our system, which is small compared to the phase-space region with quasi-periodic oscillation about the elliptic periodic orbit.
This is the many-particle analogon of non-dispersive single-particle wave packets, which were predicted and observed in microwave driven Rydberg atoms \cite{Buchleitner2002,Maeda2004,Maeda2005}. Such a wave packet retraces the periodic motion of the elliptic periodic orbit without being subjected to spreading.\\
\indent We can access a mixed phase space with regular and chaotic regions by changing the relative nonlinearity to $\Lambda=1.5$, the detuning to $\epsilon = -0.07$ and the driving strength to $A=0.07$.
For this scenario, the phase space in the absence of driving exhibits one hyperbolic and two elliptic fixed points  (Fig.\,3a, top).
The driven system shows, embedded in a large chaotic region, two regular islands originating from the two elliptic fixed points of the undriven system and two moving regular islands emerging due to a 2:1 resonance for a driving frequency $\omega=1.6\,\Omega_0$.
Here, we focus on the differences in the time evolution of states that are initially prepared close to this moving island or within the chaotic region. For fixed initial preparation $(z_0=0, \phi_0=2.51)$, we vary the time offset $t_0$ in order to start the dynamics within the chaotic region ($t_0=0.9 T$) or close to the moving island ($t_0=0.4 T$). We access two axes of the phase space by measuring particle number difference with and without a final $\pi/2$ pulse, corresponding to readout of the $z$ and the $\phi$ direction.
The undriven system shows a moderate broadening of the distribution along both axes (Fig.\,3, upper panels).
In the driven system, starting in the chaotic region we find fast broadening, eventually covering large parts of the phase space along the $z$ direction.
In contrast, by starting in the vicinity of the moving island, the distribution remains confined within the observation time and disperses even less than in the undriven system \cite{Behinaein2006}. This behavior can be intuitively grasped from the structure of the corresponding phase space  (chaotic vs. sticky region) and is in good agreement with the predictions from classical simulations. Note that in the simulations the initial state is set to $(z_0=-0.3$, $\phi_0=2.68)$, which captures loss effects during the experiment \cite{SuppInfo}.

Fig.\,4 shows the results of our investigation of the transition from regular to chaotic regions for the weakly (Fig.\,4a) and the strongly perturbed system (Fig.\,4b). In both cases, we observe a clear correspondence of the final variance  after a fixed time to the structure of the phase space at the initial preparation. In the weakly perturbed system, the variance increases in the vicinity of the hyperbolic periodic orbit and is minimal (compressed compared to the undriven case) in the regular region close to the elliptic periodic orbit. A similar behavior is found for the strongly perturbed system with a strong increase of variance in the chaotic region and a sudden drop when approaching the sticky regions close to the moving island. This drop occurs within 3.5 widths of a  minimal uncertainty state (CSS) and demonstrates the high experimental resolution of the phase space. The observations are in good agreement with the simulation (Fig.\,4, solid black line).

Our experiments reveal that the mixed phase-space structure of a driven system can be probed even in the short time dynamics of just a few driving periods by a variance analysis of the quantum states.
Furthermore, the experimental generation and verification of the Poincar\'{e}-Birkhoff scenario is a cornerstone to control the dynamics of complex many-particle systems. 
Since the phase-space structure, $\hbareff$ and decoherence processes in the system can be  precisely controlled, it is a model system for the exploration of the 
classical limit of quantum many-body systems \cite{Gardiner1997, Habib1998, Weiss2008}.

%%%%%%%%%%%%%%%%%%%%%%%%%%%%%%%%
%%%%%%%%%%%   FIG4   %%%%%%%%%%%%%%%%%
\begin{figure}[t]
    \centering{\includegraphics[scale=1]{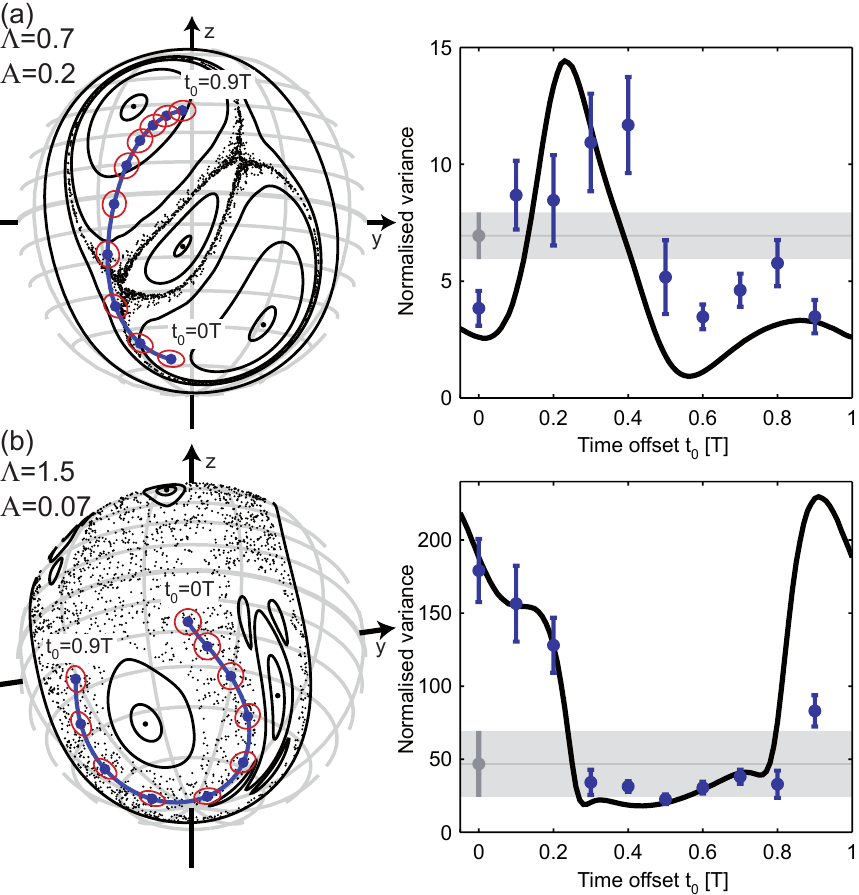} }
    \caption{(Color online).  Due to the small size of $\hbareff$  in our system, we can experimentally probe phase-space structures on small scales. The initial position is adjusted by varying the time offset $t_0$ of the driving, and is indicated in the Poincar\'{e} section for  $t_0=T$ along with the size of the effective Planck constant (left, blue points with $2 \sigma$ area in red).    
(a) Weakly perturbed system: After a fixed evolution time ($t=48$\,ms), we find  enhanced variances near the hyperbolic periodic orbit and reduced variances near the elliptic orbit. Our findings are compared to the undriven case (gray) and the classical simulation (black line). (b) Strongly perturbed regime:  The variance after an evolution time of $t=61$\,ms reveals a strong dependence on the initial preparation, indicating the high resolution in phase space.}
\end{figure}
%%%%%%%%%%%%% %%%%%%%%%%%%%%%%%%%
%%%%%%%%%%%%% %%%%%%%%%%%%%%%%%%%

\begin{acknowledgments}
We are grateful to S. Gardiner and C. Weiss for discussions and valuable feedback on the manuscript. We thank  P. Kevrekidis and L. Carr for revealing connections to general nonlinear dynamics and D. Linnemann for experimental assistance.
We acknowledge financial support through the DFG Forschergruppe 760 'Scattering Systems with Complex Dynamics' of the Deutsche Forschungsgemeinschaft, the Heidelberg Center for Quantum Dynamics and the European Comission FET-Proactive grant AQuS (Project No. 640800). W.M. acknowledges support by the Studienstiftung des deutschen Volkes. 
\end{acknowledgments}

%%%%%%%%%%%%
%Bibliography
%%%%%%%%%%%%%

\bibliographystyle{APS}

%%%%%%%%%%%%%%%%%%%%%%%%%%%%%%%

\newpage
\clearpage
%%%%%%%%%%%%
%%%%%%%%%%%%
%%%%%%%%%%%%
\onecolumngrid
\renewcommand{\figurename}{Supp.~Fig.}
\setcounter{figure}{0}   
\section*{Supplemental Material: Observing the emergence of chaos in a many-particle quantum system}
\twocolumngrid
\subsection*{Experimental system}
For our experimental studies, we employ BECs containing $N\approx 700$   $^{\text{87}}$Rb atoms in the two hyperfine levels $\ket{\downarrow} = \ket{F=1,m_F=+1}$ and $\ket{\uparrow} = \ket{F=2,m_F=-1}$. In the presence of linear coupling and collisional interaction, this system can be described with collective spin operators and is governed by the Lipkin-Meshkov-Glick Hamiltonian $\hat{H} = \chi\hat{J}_z^2-\Omega\hat{J}_x+\delta\hat{J}_z$ \cite{Zibold2010, Strobel2014, Micheli2003}. 

The  coupling is realized via a two-photon microwave and radio frequency (RF) transition with Rabi frequency $\Omega$.
A slight detuning $\delta$ of the coupling from the transition breaks the symmetry of the phase space along the $z$ direction. The coupling phase, frequency and amplitude can be controlled using an arbitrary waveform generator for the RF, and the amplitude of the microwave radiation can be changed by use of an attenuator on a two-port switch. For all experiments, AC Zeeman frequency shifts resulting from the two-photon coupling scheme and collisional mean field shifts are compensated.

An interspecies Feshbach resonance at a magnetic field of $9.1$\,G permits the enhancement of the atomic nonlinearity to $N\chi \approx 2\pi\times32(2)\,$Hz at our working field of $9.12$\,G. The resulting atom loss during the evolution time (typical time scale  $80$\,ms) causes a time-dependence of the atom number dependent collisional shifts and thus of the detuning  $\delta$. This shift is compensated by an independently calibrated feed forward on the frequency of the RF radiation. Additional temporal changes of $\chi(N)$ are not compensated, but do not change the overall structure of the phase space.
Technical fluctuations and drifts in the detuning resulting from magnetic field changes are as small as $\sigma_{\delta}\approx 2\pi\times0.45$\,Hz, as deduced from independent Ramsey measurements.

An optical lattice potential is used to simultaneously generate about 30 independent condensates, which yields enhanced statistics and freezes out  external degrees of freedom due to the high trapping potential in all spatial directions. The individual condensates are imaged using site- and state-resolved absorption imaging after Stern-Gerlach separation of the two components.

The classical limit of this system is obtained for $N\rightarrow\infty$ and keeping $\Lambda$ constant by replacing the operators with their mean values. The resulting mean field dynamics have been intensely studied both theoretically  and experimentally \cite{Smerzi1997, Zibold2010}. In this classical description, a bifurcation \cite{Zibold2010} occurs for similar strength of coupling and nonlinearity at $\Lambda_{\text{cr}}=\left( 1+|\epsilon|^{\frac 2 3} \right) ^\frac 3 2$, where one elliptic fixed point turns into one hyperbolic and two elliptic fixed points. 

\subsection*{Experimental sequence}
At the beginning of the experimental sequence, all atoms are initialized in $\ket{\downarrow}$. A high-power coupling pulse ($\Omega \approx 2\pi\times310$\,Hz) is employed to prepare the initial quantum state at a fixed initial population imbalance $z_0$ by properly adjusting the length of the pulse.

Subsequently, the power of microwave and RF radiation is attenuated to reach the desired value of $\Lambda$, and the phase of the RF radiation is nonadiabatically adjusted to prepare the initial state at the chosen phase $\phi_0$. The driving is realized via the corresponding periodic modulation of the RF signal amplitude. After  the evolution time $t$, the two components are absorption imaged to obtain the $z$ projection of the final state. Alternatively, for readout along the  $y = \sin(\phi)$ direction, we apply a further fast $\pi/2$ rotation around the $x$ direction, which converts the phase information into  population imbalance.

\subsection*{Data analysis}
The results are postselected on atom number following the actual loss rate in order to ensure  a consistent ensemble starting with $N=700 \pm 25$ at $t=0$. Each data point comprises 60 to 90 independent realizations of the experiment.
Outliers are rejected by a modified Z-score test.

  The obtained variances in $z$ are normalized by $\text{Var}_{\text{CSS}} =  (1-\langle z \rangle^2 )\times N(t)$, where $N(t)$ denotes the total atom number at time $t$. This ensures that the state  is referenced to the corresponding CSS regardless of its position on the sphere or evolution time in the presence of atom loss.
 The error bar of the variance is given by the 1 s.d. confidence interval obtained from Jackknife resampling.
 
\subsection*{Simulations}
For the comparison of the many-particle quantum system with the classical model (Eq.~1 of the main text), the finite width of the wave function is accounted for by integrating the corresponding equations of motion for a set of points sampled from the Gaussian distribution of the corresponding coherent spin state. From these integrations, the corresponding mean values and variances are obtained.
While the change of the collisional shifts due to atom loss is experimentally compensated, the corresponding change of the relative nonlinearity $\Lambda$ is not corrected. To obtain a proper phase space description, this time dependence is captured in the simulations by using an effective starting point, which is obtained from fits for both parameter sets from reference measurements with the undriven system. \\

%%%%%%%%%%%%
%Bibliography
%%%%%%%%%%%%%

\bibliographystyle{APS}

\begin{thebibliography}{10}
\expandafter\ifx\csname url\endcsname\relax
  \def\url#1{\texttt{#1}}\fi
\expandafter\ifx\csname urlprefix\endcsname\relax\def\urlprefix{URL }\fi
\providecommand{\bibinfo}[2]{#2}
\providecommand{\eprint}[2][]{\url{#2}}
%
% 
% 

\bibitem{Birkhoff1927}
\bibinfo{author}{G.~D.~Birkhoff,}
\newblock \textit{{\bibinfo{title}{On the periodic motions of dynamical systems.}}}
\newblock {\bibinfo{journal}{Acta Mathematica}}
\textbf{\bibinfo{volume}{50}},
\bibinfo{pages}{359--379}
(\bibinfo{year}{1927}).

\bibitem{Ott2002}
\bibinfo{author}{E.~Ott,} 
\newblock  \textit{\bibinfo{title}{{Chaos in dynamical systems.}}}
 (\bibinfo{publisher}{Cambridge  Univ. Press, Cambridge}, \bibinfo{year}{2002}).
 
 \bibitem{Weiss2008}
\bibinfo{author}{C.~Weiss} and
\bibinfo{author}{N.~Teichmann,}
\newblock \textit{{\bibinfo{title}{Differences between Mean-Field Dynamics and N-Particle Quantum Dynamics as a Signature of Entanglement.}}}
\newblock  {\bibinfo{journal}{Phys. Rev. Lett.}}
\textbf{\bibinfo{volume}{100}},
\bibinfo{pages}{140408}
(\bibinfo{year}{2008}).



\bibitem{Xie2005}
\bibinfo{author}{Q.~Xie} and
\bibinfo{author}{W.~Hai},
\newblock \textit{{\bibinfo{title}{Quantum entanglement and chaos in kicked two-component Bose-Einstein condensates.}}}
\newblock  {\bibinfo{journal}{Eur. Phys. J. D}}
\textbf{\bibinfo{volume}{33}},
\bibinfo{pages}{265--272}
(\bibinfo{year}{2005}).

\bibitem{Ghose2008}
\bibinfo{author}{S.~Ghose},
\bibinfo{author}{R.~Stock},
\bibinfo{author}{P.~Jessen},
\bibinfo{author}{R.~Lal}, and
\bibinfo{author}{A.~Silberfarb},
\newblock \textit{{\bibinfo{title}{Chaos, entanglement, and decoherence in the quantum kicked top.}}}
\newblock  {\bibinfo{journal}{Phys. Rev. A}}
\textbf{\bibinfo{volume}{78}},
\bibinfo{pages}{042318}
(\bibinfo{year}{2008}).

\bibitem{Lin1990a}
\bibinfo{author}{W.~A.~Lin} and
\bibinfo{author}{L.~E.~Ballentine},
\newblock \textit{{\bibinfo{title}{Quantum Tunneling and Chaos in a Driven Anharmonic Oscillator.}}}
\newblock  {\bibinfo{journal}{Phys. Rev. Lett.}}
\textbf{\bibinfo{volume}{65}},
\bibinfo{pages}{2927--2930}
(\bibinfo{year}{1990}).


\bibitem{Tomsovic1994}
\bibinfo{author}{S.~Tomsovic} and
\bibinfo{author}{D.~Ullmo},
\newblock \textit{{\bibinfo{title}{Chaos-assisted tunneling.}}}
\newblock  {\bibinfo{journal}{Phys. Rev. E}}
\textbf{\bibinfo{volume}{50}},
\bibinfo{pages}{145--162}
(\bibinfo{year}{1994}).

\bibitem{Korsch2008CAT}
\bibinfo{author}{M.~P. Strzys},
\bibinfo{author}{E.~M.~Graefe}, and
\bibinfo{author}{H.~J.~Korsch},
\newblock \textit{{\bibinfo{title}{Kicked {B}ose--{H}ubbard systems and kicked tops -- destruction and stimulation of tunneling.}}}
\newblock  {\bibinfo{journal}{New J. Phys.}}
\textbf{\bibinfo{volume}{10}},
\bibinfo{pages}{013024}
(\bibinfo{year}{2008}).
 
\bibitem{Moore1994}
\bibinfo{author}{F.~L. Moore},
\bibinfo{author}{J.~C. Robinson},
\bibinfo{author}{C. Bharucha},
\bibinfo{author}{P.~E. Williams}, and
\bibinfo{author}{M.~G.~Raizen},
\newblock \textit{{\bibinfo{title}{Observation of Dynamical Localization in Atomic Momentum Transfer: A New Testing Ground for Quantum Chaos.}}}
\newblock  {\bibinfo{journal}{Phys. Rev. Lett.}}
\textbf{\bibinfo{volume}{73}},
\bibinfo{pages}{2974--2977}
(\bibinfo{year}{1994}).

\bibitem{Moore1995}
\bibinfo{author}{F.~L.~Moore},
\bibinfo{author}{J.~C.~Robinson},
\bibinfo{author}{C.~F.~Bharucha},
\bibinfo{author}{B.~Sundaram}, and
\bibinfo{author}{M.~G.~Raizen},
\newblock  \textit{{\bibinfo{title}{Atom Optics Realization of the Quantum $\delta{}$-Kicked Rotor.}}}
\newblock {\bibinfo{journal}{Phys. Rev. Lett.}}
\textbf{\bibinfo{volume}{75}},
\bibinfo{pages}{4598--4601}
(\bibinfo{year}{1995}).

\bibitem{Steck2001}
\bibinfo{author}{D.~A.~Steck},
\bibinfo{author}{W.~H.~Oskay}, and
\bibinfo{author}{M.~G.~Raizen},
\newblock \textit{{\bibinfo{title}{Observation of Chaos-Assisted Tunneling Between Islands of Stability.}}}
\newblock  {\bibinfo{journal}{Science}}
\textbf{\bibinfo{volume}{293}},
\bibinfo{pages}{274--278}
(\bibinfo{year}{2001}).

\bibitem{Hensinger2001}
\bibinfo{author}{W.~K.~Hensinger}{~et~al.},
\newblock \textit{{\bibinfo{title}{Dynamical tunnelling of ultracold atoms.}}}
\newblock  {\bibinfo{journal}{Nature}}
\textbf{\bibinfo{volume}{412}},
\bibinfo{pages}{52--55}
(\bibinfo{year}{2001}).

\bibitem{Chaudhury2009}
\bibinfo{author}{S.~Chaudhury},
\bibinfo{author}{A.~Smith},
\bibinfo{author}{B.~E.~Anderson},
\bibinfo{author}{S.~Ghose}, and
\bibinfo{author}{P.~S.~Jessen},
\newblock \textit{{\bibinfo{title}{Quantum signatures of chaos in a kicked top.}}}
\newblock  {\bibinfo{journal}{Nature}}
\textbf{\bibinfo{volume}{461}},
\bibinfo{pages}{768--771}
(\bibinfo{year}{2009}).


\bibitem{Duffy2004}
\bibinfo{author}{G.~J.~Duffy},
\bibinfo{author}{S.~Parkins},
\bibinfo{author}{T.~M\"{u}ller},
\bibinfo{author}{M.~Sadgrove},
\bibinfo{author}{R.~Leonhardt}, and
\bibinfo{author}{A.~C.~Wilson},
\newblock \textit{{\bibinfo{title}{Experimental investigation of early-time diffusion in the quantum kicked rotor using a Bose-Einstein condensate.}}}
\newblock  {\bibinfo{journal}{Phys. Rev. E}}
\textbf{\bibinfo{volume}{70}},
\bibinfo{pages}{056206}
(\bibinfo{year}{2004}).



\bibitem{Behinaein2006}
\bibinfo{author}{G.~Behinaein},
\bibinfo{author}{V.~Ramareddy},
\bibinfo{author}{P.~Ahmadi}, and
\bibinfo{author}{G.~S.~Summy},
\newblock \textit{{\bibinfo{title}{Exploring the Phase Space of the Quantum $\delta$-Kicked accelerator.}}}
\newblock  {\bibinfo{journal}{Phys. Rev. Lett.}}
\textbf{\bibinfo{volume}{97}},
\bibinfo{pages}{244101}
(\bibinfo{year}{2006}).


\bibitem{Frahm1985}
\bibinfo{author}{H.~Frahm} and
\bibinfo{author}{H.~J.~Mikeska},
\newblock \textit{{\bibinfo{title}{On the Dynamics of a Quantum System which is Classically Chaotic.}}}
\newblock  {\bibinfo{journal}{Z. Phys. B Condensed Matter}}
\textbf{\bibinfo{volume}{60}},
\bibinfo{pages}{117--126}
(\bibinfo{year}{1985}).

\bibitem{Salmond2002}
\bibinfo{author}{G.~L.~Salmond},
\bibinfo{author}{C.~A.~Holmes}, and
\bibinfo{author}{G.~J.~Milburn},
\newblock \textit{{\bibinfo{title}{Dynamics of a strongly driven two-component {B}ose-{E}instein condensate.}}}
\newblock  {\bibinfo{journal}{Phys. Rev. A}}
\textbf{\bibinfo{volume}{65}},
\bibinfo{pages}{033623}
(\bibinfo{year}{2002}).



\bibitem{SuppInfo}
\bibinfo{title}{See Supplemental Material for details on the experimental sequence, data analysis and simulations.}

\bibitem{Micheli2003}
\bibinfo{author}{A.~Micheli}, 
\bibinfo{author}{D.~Jaksch},
\bibinfo{author}{J.~I.~Cirac}, and
\bibinfo{author}{P.~Zoller},
\newblock  \textit{{\bibinfo{title}{Many-particle entanglement in two-component Bose-Einstein condensates.}}}
\newblock {\bibinfo{journal}{Phys. Rev. A}}
\textbf{\bibinfo{volume}{67}},
\bibinfo{pages}{013607}
(\bibinfo{year}{2003}).

\bibitem{Zibold2010}
\bibinfo{author}{T.~Zibold},
\bibinfo{author}{E.~Nicklas},
\bibinfo{author}{C.~Gross}, and
\bibinfo{author}{M.~K.~Oberthaler},
\newblock \textit{{\bibinfo{title}{Classical Bifurcation at the Transition from {R}abi to {J}osephson Dynamics.}}}
\newblock  {\bibinfo{journal}{Phys. Rev. Lett.}}
\textbf{\bibinfo{volume}{105}},
\bibinfo{pages}{204101}
(\bibinfo{year}{2010}).

\bibitem{Strobel2014}
\bibinfo{author}{H.~Strobel}~et~al.,
\newblock \textit{{\bibinfo{title}{ Fisher information and entanglement of non-Gaussian spin states.}}}
\newblock  {\bibinfo{journal}{Science}}
\textbf{\bibinfo{volume}{345}},
\bibinfo{pages}{424--427}
(\bibinfo{year}{2014}).



\bibitem{Gardiner2000}
\bibinfo{author}{S.~A.~Gardiner},
\bibinfo{author}{D.~Jaksch},
\bibinfo{author}{R.~Dum},
\bibinfo{author}{J.~I.~Cirac}, and
\bibinfo{author}{P.~Zoller},
\newblock \textit{{\bibinfo{title}{Nonlinear matter wave dynamics with a chaotic potential.}}}
\newblock  {\bibinfo{journal}{Phys. Rev. A}}
\textbf{\bibinfo{volume}{62}},
\bibinfo{pages}{023612}
(\bibinfo{year}{2000}).



\bibitem{Monteiro2008}
\bibinfo{author}{T.~S.~Monteiro},
\bibinfo{author}{A.~Ran\c{c}on}, and
\bibinfo{author}{J.~Ruostekoski},
\newblock \textit{{\bibinfo{title}{Nonlinear Resonances in $\delta$-Kicked Bose-Einstein Condensates.}}}
\newblock  {\bibinfo{journal}{Phys. Rev. Lett.}}
\textbf{\bibinfo{volume}{102}},
\bibinfo{pages}{014102}
(\bibinfo{year}{2009}).

\bibitem{Smerzi1997}
\bibinfo{author}{A.~Smerzi},
\bibinfo{author}{S.~Fantoni},
\bibinfo{author}{S.~Giovanazzi}, and
\bibinfo{author}{S.~R.~Shenoy},
\newblock \textit{{\bibinfo{title}{Quantum Coherent Atomic Tunneling between Two Trapped {B}ose-{E}instein Condensates.}}}
\newblock  {\bibinfo{journal}{Phys. Rev. Lett.}}
\textbf{\bibinfo{volume}{79}},
\bibinfo{pages}{4950--4953}
(\bibinfo{year}{1997}).

\bibitem{Kronjaeger2008}
\bibinfo{author}{J.~Kronj\"{a}ger},
\bibinfo{author}{K.~Sengstock}, and
\bibinfo{author}{K.~Bongs},
\newblock \textit{{\bibinfo{title}{Chaotic dynamics in spinor Bose-Einstein
condensates.}}}
\newblock  {\bibinfo{journal}{New J. Phys.}}
\textbf{\bibinfo{volume}{10}},
\bibinfo{pages}{045028}
(\bibinfo{year}{2008}).

\bibitem{Cheng2010}
\bibinfo{author}{J.~Cheng},
\newblock \textit{{\bibinfo{title}{Chaotic dynamics in a periodically driven spin-1 condensate.}}}
\newblock  {\bibinfo{journal}{Phys. Rev. A}}
\textbf{\bibinfo{volume}{81}},
\bibinfo{pages}{023619}
(\bibinfo{year}{2010}).


\bibitem{Buchleitner2002}
\bibinfo{author}{A.~Buchleitner},
\bibinfo{author}{D.~Delande}, and
\bibinfo{author}{J.~Zakrzewski},
\newblock \textit{{\bibinfo{title}{Non-dispersive wave packets in periodically driven quantum systems.}}}
\newblock  {\bibinfo{journal}{Physics Reports}}
\textbf{\bibinfo{volume}{368}},
\bibinfo{pages}{409--547}
(\bibinfo{year}{2002}).

\bibitem{Maeda2004}
\bibinfo{author}{H. Maeda} and
\bibinfo{author}{T.~F.~Gallagher},
\newblock \textit{{\bibinfo{title}{Nondispersing Wave Packets.}}}
\newblock  {\bibinfo{journal}{Phys. Rev. Lett.}}
\textbf{\bibinfo{volume}{92}},
\bibinfo{pages}{133004}
(\bibinfo{year}{2004}).

\bibitem{Maeda2005}
\bibinfo{author}{H. Maeda},
\bibinfo{author}{D.~V.~L. Norum}, and
\bibinfo{author}{T.~F.~Gallagher},
\newblock \textit{{\bibinfo{title}{Microwave Manipulation of an Atomic Electron in a Classical Orbit.}}}
\newblock  {\bibinfo{journal}{Science}}
\textbf{\bibinfo{volume}{307}},
\bibinfo{pages}{1757--1760}
(\bibinfo{year}{2005}).


\bibitem{Gardiner1997}
\bibinfo{author}{S.~A.~Gardiner},
\bibinfo{author}{J.~I.~Cirac,} and
\bibinfo{author}{P.~Zoller,}
\newblock \textit{{\bibinfo{title}{Quantum Chaos in an Ion Trap: The Delta-Kicked Harmonic Oscillator.}}}
\newblock  {\bibinfo{journal}{Phys. Rev. Lett.}}
\textbf{\bibinfo{volume}{79}},
\bibinfo{pages}{4790-4793}
(\bibinfo{year}{1997}).


\bibitem{Habib1998}
\bibinfo{author}{S.~Habib},
\bibinfo{author}{K.~Shizume}, and
\bibinfo{author}{W.~H.~Zurek},
\newblock \textit{{\bibinfo{title}{Decoherence, Chaos, and the Correspondence Principle.}}}
\newblock  {\bibinfo{journal}{Phys. Rev. Lett.}}
\textbf{\bibinfo{volume}{80}},
\bibinfo{pages}{4361-4365}
(\bibinfo{year}{1998}).





\end{thebibliography}

\begin{thebibliography}{10}
\expandafter\ifx\csname url\endcsname\relax
  \def\url#1{\texttt{#1}}\fi
\expandafter\ifx\csname urlprefix\endcsname\relax\def\urlprefix{URL }\fi
\providecommand{\bibinfo}[2]{#2}
\providecommand{\eprint}[2][]{\url{#2}}
%

\bibitem{Micheli2003}
\bibinfo{author}{A.~Micheli}, 
\bibinfo{author}{D.~Jaksch},
\bibinfo{author}{J.~I.~Cirac}, and
\bibinfo{author}{P.~Zoller},
\newblock  \textit{{\bibinfo{title}{Many-particle entanglement in two-component Bose-Einstein condensates.}}}
\newblock {\bibinfo{journal}{Phys. Rev. A}}
\textbf{\bibinfo{volume}{67}},
\bibinfo{pages}{013607}
(\bibinfo{year}{2003}).

\bibitem{Zibold2010}
\bibinfo{author}{T.~Zibold},
\bibinfo{author}{E.~Nicklas},
\bibinfo{author}{C.~Gross}, and
\bibinfo{author}{M.~K.~Oberthaler},
\newblock \textit{{\bibinfo{title}{Classical Bifurcation at the Transition from {R}abi to {J}osephson Dynamics.}}}
\newblock  {\bibinfo{journal}{Phys. Rev. Lett.}}
\textbf{\bibinfo{volume}{105}},
\bibinfo{pages}{204101}
(\bibinfo{year}{2010}).

\bibitem{Strobel2014}
\bibinfo{author}{H.~Strobel}~et~al.,
\newblock \textit{{\bibinfo{title}{ Fisher information and entanglement of non-Gaussian spin states.}}}
\newblock  {\bibinfo{journal}{Science}}
\textbf{\bibinfo{volume}{345}},
\bibinfo{pages}{424--427}
(\bibinfo{year}{2014}).

\bibitem{Smerzi1997}
\bibinfo{author}{A.~Smerzi},
\bibinfo{author}{S.~Fantoni},
\bibinfo{author}{S.~Giovanazzi}, and
\bibinfo{author}{S.~R.~Shenoy},
\newblock \textit{{\bibinfo{title}{Quantum Coherent Atomic Tunneling between Two Trapped {B}ose-{E}instein Condensates.}}}
\newblock  {\bibinfo{journal}{Phys. Rev. Lett.}}
\textbf{\bibinfo{volume}{79}},
\bibinfo{pages}{4950--4953}
(\bibinfo{year}{1997}).

\end{thebibliography}

\end{document}